\documentclass[10pt]{iopart}

\usepackage{graphicx,bbm,amssymb,amsopn,braket,iopams}
\usepackage[top=1.25 in, bottom=1.25 in, left=0.9 in, right=0.9 in]{geometry}
\usepackage[labelfont={normal,bf}, font={it}]{caption}
\usepackage{color}

\def\one{\mathbbm{1}}
\def\mn{\scalebox{0.75}{-}}

\def\End{{\,{\rm End}\,}}

\newenvironment{bew}[2]{\removelastskip\vspace{6pt}\noindent
 {\it Proof  #1.}~\rm#2}{\par\vspace{6pt}}

\begin{document}
\title[Quasilocal Conservation Laws in the Quantum Hirota Model]{Quasilocal Conservation Laws in the Quantum Hirota Model}
\author{Lenart Zadnik and Toma\v{z} Prosen}
\address{Physics Department, Faculty of Mathematics and Physics, University of Ljubljana, Jadranska 19, SI-1000 Ljubljana, Slovenia}

\begin{abstract}
Extensivity of conservation laws of the quantum Hirota model on a $1+1$ dimensional lattice is considered. This model can be interpreted in terms of an integrable many-body quantum Floquet dynamics. We establish the procedure to generate a continuous family of quasilocal conservation laws from the conserved operators proposed by Faddeev and Volkov. The Hilbert-Schmidt kernel which allows the calculation of inner products of these new conservation laws is explicitly computed. This result has potential applications in quantum quench and transport problems in integrable quantum field theories.
\end{abstract}

\section{Introduction}
In recent years, the study of integrable systems out of equilibrium has become one of the main focuses of theoretical and mathematical physics \cite{jstat}.
In particular, understanding the local and also quasilocal\footnote{A weaker version of locality, that is explained later.} conservation laws and their impact on the non-equilibrium dynamics of integrable systems has become an important problem of  quantum statistical physics (see, e.g., the recent review \cite{review} and references therein). Apart from their importance in the problem of local equilibration of isolated systems towards the Generalized Gibbs Ensemble, integrals of motion prove to be useful also in the linear response theory of the transport phenomena. Notably, conservation laws varying linearly in the system size, as measured by the Hilbert-Schmidt norm, can be used in the Mazur-Suzuki bound to  rigorously establish the ballistic transport at high temperatures \cite{P11}. Until now these ideas have been implemented mainly in the paradigmatic example of integrable systems, the spin-$1/2$ $XXZ$ model. 

The notion of the quasilocal conservation laws has so far only been studied in the lattice models with finite dimensional local Hilbert spaces, such as spin chains. There has been an alternative proposal of quasilocal charges in continuous field theories \cite{panfil}, but it seems that this approach can only be worked out explicitly for the free theories.  

Our proposition here is to study quasilocality in the integrable lattice regularization of an interacting field theory in $1+1$ dimensions, namely the quantum sine-Gordon (SG) model. 
In particular we shall consider the so-called quantum Hirota model, put forward by Faddeev and Volkov \cite{faddeev} (see also \cite{faddeev2}) which is, in our opinion, the most elegant and clean lattice regularization of the SG model. In contrast to the fermionic light-cone lattice approach of Destri and De Vega \cite{destri}, the Hirota model uses (multiplicative) bosonic variables. 
If one interprets the space-time lattice as a two-dimensional lattice of discretized spectral and spin parameters, then the quantum Hirota model becomes equivalent to the T-system describing a fusion hierarchy of transfer matrices (see, e.g., \cite{review} and references therein).

In this paper, the quasilocal conservation laws of the quantum Hirota model are identified for a generic root-of-unity quantization parameter, where the local Hilbert space is finite dimensional. We build on the seminal results of Faddeev and Volkov \cite{faddeev}, where integrability of this lattice model has been established and the transfer matrix constructed. The quantum Hirota model, a version of which is also known under the name of quantum Volterra model \cite{volkov} can be interpreted in multiple ways. As already discussed, one can think of it as a light-like lattice regularization of the SG quantum field theory \cite{faddeev}, describing for example the low energy physics of the anisotropic Heisenberg model, or as a quantized Volterra model, the classical counterpart of which is used in the study of population dynamics \cite{volkov}.  It is also closely related to the Chiral Potts model \cite{bazhanov2}, i.e., a classical statistical model with discrete cyclic $\mathbbm{Z}_m$ variables on a 2D lattice. However, our favorite interpretation of this model is in terms of a Floquet (periodically) driven system with discrete cyclic (Weyl) variables  --- a quantum protocol interchangeably propagating dynamical variables at even and odd lattice sites that is completely determined by a local recursive dynamical rule \`{a} la quantum cellular automaton.

For the lattice systems it is convenient to speak of a linear extensivity in the sense of the Hilbert-Schmidt (HS) inner product. Let $A$, $B$ and $\mathbbm{1}$ be the operators on a Hilbert space, $\mathbbm{1}$ denoting the identity. Then one defines the HS product of operators and the corresponding norm as\footnote{Note that this inner product is semi-definite since, in addition to $0$, all operators of the form $\alpha \mathbbm{1}$ with $\alpha\in\mathbbm{C}$ also have HS norm equal to $0$.}
\begin{equation}
\langle A,B\rangle=\frac{\tr{\left(A^\dagger B\right)}}{\tr{\mathbbm{1}}}-\frac{\tr{(A^\dagger)}}{\tr{\mathbbm{1}}}\frac{\tr{(B)}}{\tr{\mathbbm{1}}},\qquad \|A\|^2_{\rm HS}=\langle A,A\rangle .\label{def_hs}
\end{equation}
Linear extensivity of an observable $Q$ acting on the full Hilbert space is then just
\begin{equation}
\|Q\|^2_{\rm HS}\propto N,
\end{equation}
where $N$ denotes the number of lattice sites, i.e., the system size. The most commonly known linearly extensive operators are local operators, that is, translationally invariant sums $\sum_j h_j^{[r]}$ of the local operator-valued densities $h_{j}^{[r]}$ acting nontrivially on clusters of $r$ adjacent lattice sites, starting at the site $j$. Here $r$ is fixed while the sum goes over all the lattice sites $j$. As an example we can take the Heisenberg Hamiltonian, where $r=2$. A non-local operator can still be linearly extensive if it satisfies the condition of quasilocality, more specifically, if it is a double sum of local densities $\sum_j \sum_r h_{j}^{[r]}$, where also $r$ is allowed to change, provided that these densities obey
\begin{equation}
\| h_{j}^{[r]}\|^2_{\rm HS} \le C e^{-\gamma r},
\end{equation} 
for some $C,\gamma > 0$. Our aim is to construct linearly extensive conservation laws for the Hirota model, starting from the integrals of motion constructed by Faddeev and Volkov \cite{faddeev}, using the procedure developed for the isotropic Heisenberg model in \cite{IMP15}. We should stress that, by themselves, the conserved quantities of Faddeev and Volkov are not linearly extensive in the sense of the HS norm.

The main result presented in sections \ref{QL} and \ref{Kernel} of the paper can be summarized as follows. Let us write the root-of-unity quantization parameter of the Hirota model as $q=\exp(i\frac{\ell}{m}\pi)$, $m$ being an odd integer ($\ell<m$, $\ell$ even) and denote $\Lambda_s(\lambda)=1+(\kappa^2+\kappa^{-2})\lambda^2+\lambda^4$, where $\kappa$ is the scaling parameter and $\lambda$ a complex number (see Section~\ref{Dynamics} for the details on how these parameters enter the discussion). Additionally let $T(\lambda)$ be the commuting transfer matrix of the quantum Hirota model as proposed by Faddeev and Volkov in \cite{faddeev, volkov}. The 
conserved charge
\begin{equation}
X(\lambda)=\frac{1}{\Lambda_s(\lambda)^N}\,T(\lambda q^{\frac{1}{2}})\frac{\rmd}{\rmd\lambda}T(\lambda q^{-\frac{1}{2}}),\label{result1}
\end{equation}
is a quasilocal operator for $\lambda\in\mathbbm{C}\setminus\{0\}$ with $\arg{\lambda}\in(\pi-\eta\frac{\pi}{2m},\pi+\eta\frac{\pi}{2m})\cup(-\eta\frac{\pi}{2m},\eta\frac{\pi}{2m})$ where $\eta=\min(\ell,m-\ell)$. Moreover, in the thermodynamic limit, linear extensivity holds for this conservation law since
\begin{equation}
\langle X(\lambda),X(\mu)\rangle=N\,\mathcal{K}(\lambda,\mu)+\mathcal{O}(e^{-\gamma N}),\qquad \gamma>0,
\end{equation}
where the Hilbert-Schmidt kernel $\mathcal{K}(\lambda,\mu)$ is explicitly computed. Quasilocality of \eref{result1} is proven analytically for a general root of unity, except for the precise conditions on the domain of the spectral parameter $\lambda$. The latter is deduced from the results of an exact numerical diagonalization.

In Section \ref{Dynamics}, the definition of the Hirota model will be revisited along with its dynamics, while Section \ref{Integrals} describes the Faddeev-Volkov \cite{faddeev, volkov} conservation laws, constructed as a part of the algebraic Bethe ansatz approach. The last two sections, \ref{QL} and \ref{Kernel}, constitute the explanation of the results -- in Section \ref{QL} linear extensivity of charges \eref{result1} is established, following the procedure proposed in \cite{IMP15}, while in Section \ref{Kernel} the Hilbert-Schmidt kernel is explicitly computed and an explicit matrix product form of the conserved charges is spelled out. Some of the technical details are summed up in the appendices, along with an example of a Floquet interpretation of the model.

\section{The dynamics of the quantum Hirota model}\label{Dynamics}

\begin{figure}[h]
\includegraphics[width=450pt]{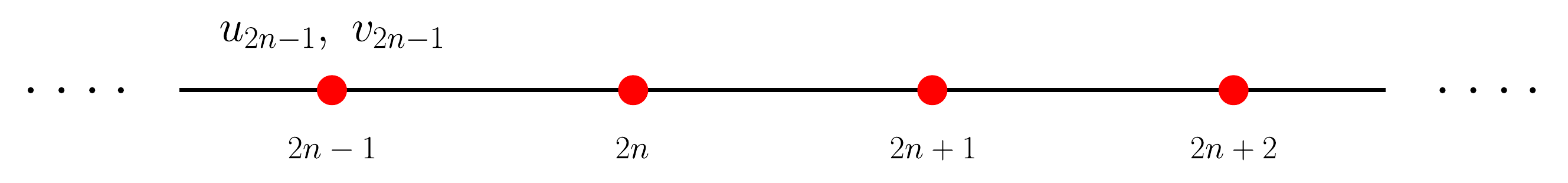}
\caption{Each physical lattice site is equipped with an action of the two Weyl variables $u,v$ that satisfy the $q$-deformed Weyl algebra. Weyl variables belonging to different sites commute.}
\label{fig:chain_0}
\end{figure}

\begin{figure}[h]
\includegraphics[width=450pt]{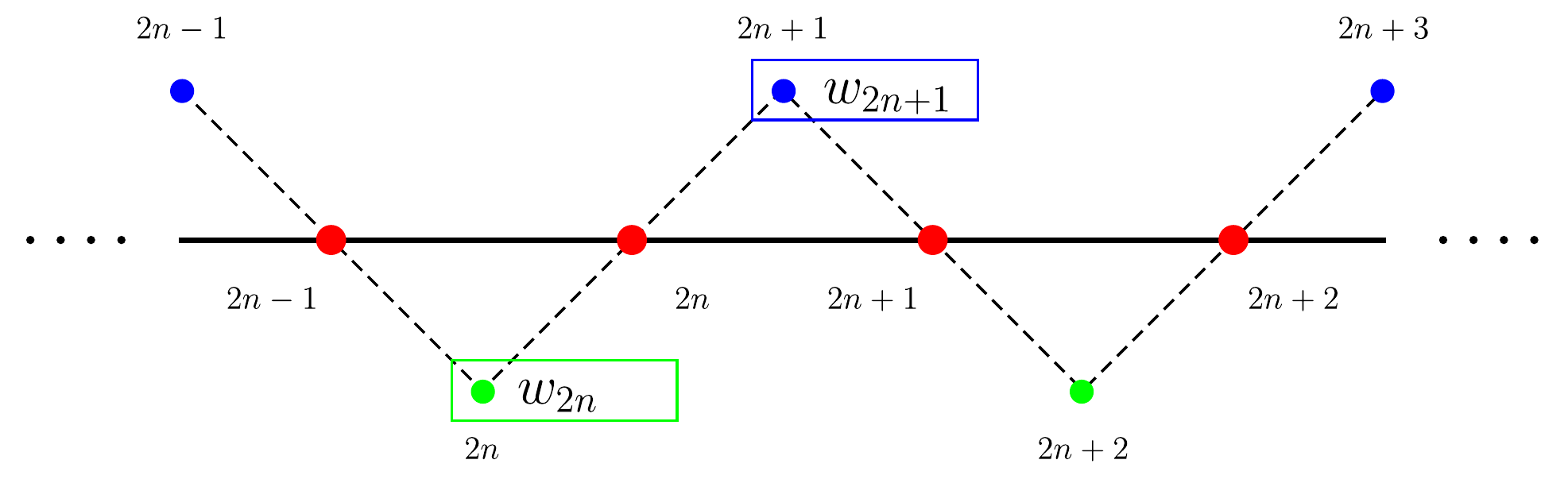}
\caption{Layout of the dynamical variables. The odd-numbered dynamical variables are denoted by the blue dots, while the even-numbered are green.}
\label{fig:chain_1}
\end{figure}

Consider a periodic chain of $2N$ sites, where each site corresponds to a local \emph{physical} Hilbert space $\mathcal{H}$ acted upon by a pair of Weyl variables $u,v\in\End(\mathcal{H})$. These satisfy the $q$-deformed canonical commutation relation,
\begin{equation}
uv=qvu,\label{weyl}
\end{equation}
where complex number $q$ is a root of unity, $q^m=1$, and $m$ an odd integer. For example, in the case $m=3$ we have -- up to a similarity transformation -- a unique unitary matrix representation on a $3-$dimensional physical Hilbert space $\mathcal{H}$, of the form
\begin{equation}
u=
\pmatrix{0 & 1 & 0\cr
0 & 0 & 1\cr
1 & 0 & 0},\qquad
v=
\pmatrix{
1 & 0 & 0\cr
0 & q & 0\cr
0 & 0 & q^2},\qquad q=e^{i\,2\pi/3}.
\end{equation}
Such representations can be constructed in a similar way for any $m$.
Using the matrix representation, the complete set of Weyl variables is given in terms of the tensor products
$u_j = \one_{m^{j-1}} \otimes u\otimes \one_{m^{2N-j}}$,
$v_j = \one_{m^{j-1}}\otimes v\otimes \one_{m^{2N-j}}$,
so that $u_j v_k = v_k u_j$ for $k\neq j$. Here $\mathbbm{1}_d$ denotes a $d\times d$ identity matrix. These tensor products act on the full physical Hilbert space $\mathcal{H}^{\otimes 2N}$ of the system.
The basic setting is shown in figure~\ref{fig:chain_0}.
In order to define the dynamical evolution, we imagine a \emph{zigzag} chain, intertwining with our physical chain. This zigzag chain is equipped with the {\em dynamical variables} $w_j=u_{j-1}v_{j-1}u_jv^{-1}_j$, as shown in figure~\ref{fig:chain_1}.
For the neighbouring dynamical variables $w_{j-1},w_{j}$, the following  algebraic relations hold:
\begin{equation}
w_{j-1}w_{j}=q^2 w_{j}w_{j-1},\qquad j=1,2,...,2N.
\end{equation}
Here we have taken $w_0=w_{2N}$ due to the periodicity of the chain. Non-neighbouring variables $w_j$ commute, since they have no physical operators in common. Note in particular, that the odd-numbered and the even-numbered variables commute among themselves, respectively.

\begin{figure}
\includegraphics[width=400pt]{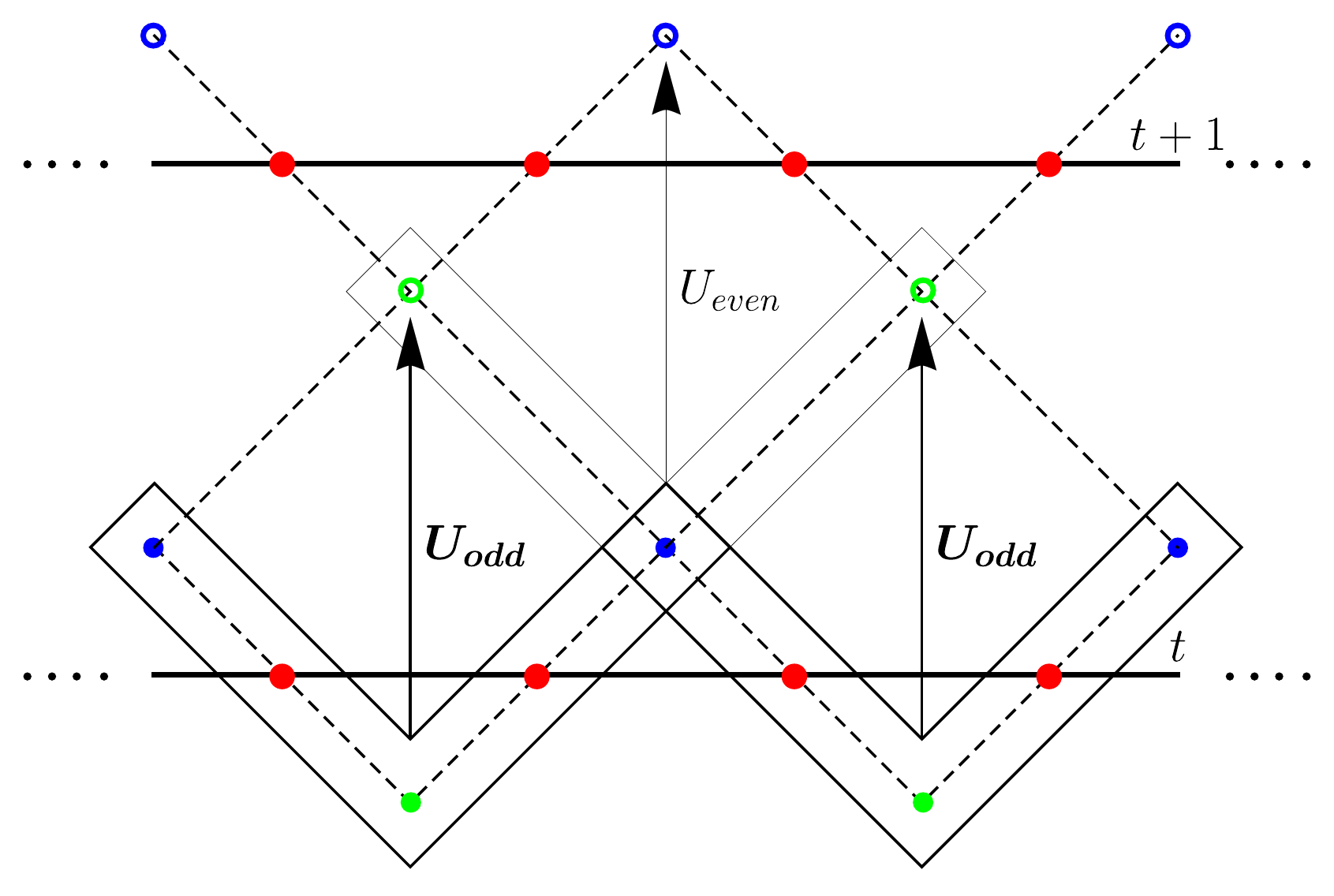}
\caption{The propagation corresponds to first kicking the even degrees of freedom for one time step with $U_{\rm odd}$ and then using them to kick the odd degrees of freedom for the same time step with $U_{\rm even}$. Empty circles represent variables propagated for the full time step. We thus consider the quantum protocol which transforms the blue (green) filled circles into the blue (green) empty circles, while using the physical variables (filled red circles) as a convenient static operator basis.}
\label{fig:propagation}
\end{figure}

Suppose the discrete time evolution $U:\,\mathbbm{N}_0\longrightarrow\End(\mathcal{H}^{\otimes 2N})$ can be factorized as 
\begin{equation}
U(t)=U_{\rm even}(t)U_{\rm odd}(t)=\prod_{n=1}^N r(\kappa^2,w_{2n}(t))\prod_{m=1}^N r(\kappa^2,w_{2m-1}(t)),\label{propagator}
\end{equation}
$r$ being some analytic function of the dynamical variable $w$ and the square of the \emph{scaling} parameter $\kappa$.
We have denoted our time slice by $t\in\mathbbm{N}_0$. Let us propagate $w_{2n}(t)$ according to $w_{2n}(t+1)=U^{-1}(t)\,w_{2n}(t)\,U(t)$, which amounts to
\begin{equation}
\fl w_{2n}(t+1)=\left[r(\kappa^2,q^2w_{2n+1}(t))r(\kappa^2,w_{2n+1}(t))^{-1}\right]  w_{2n}(t)\,\left[r(\kappa^2,q^2w_{2n-1}(t))^{-1}r(\kappa^2,w_{2n-1}(t))\right].
\end{equation}
Propagation of the odd-numbered dynamical variables is a bit different since the even-numbered variables are already time shifted. It corresponds to
\begin{equation*}
\fl w_{2n+1}(t+1)=\left[r(\kappa^2,q^2w_{2n+2}(t+1))r(\kappa^2,w_{2n+2}(t+1))^{-1}\right]
 w_{2n+1}(t)\,\left[r(\kappa^2,q^2w_{2n}(t+1))^{-1}r(\kappa^2,w_{2n}(t+1))\right]. 
\end{equation*}
Schematically, the local propagation is shown in figure~\ref{fig:propagation}.
If we demand that the function $r$ solves the following functional equation
\begin{equation}
\frac{r(\kappa^2,q w)}{r(\kappa^2,q^{-1}w)}=f(w),\qquad f(w)=\frac{1+\kappa^2 w}{\kappa^2+w},\label{r_condition}
\end{equation}
we can rewrite the dynamical map as
\eqnarray
&w_{2n}(t+1)=f(qw_{2n+1}(t))\,w_{2n}(t)\,f(q w_{2n-1}(t))^{-1},\\[1em]
&w_{2n+1}(t+1)=f(qw_{2n+2}(t+1))\,w_{2n+1}(t)\,f(q w_{2n}(t+1))^{-1},
\endeqnarray
which in a certain continuum limit defines the dynamics of the sine-Gordon model \cite{faddeev}. There $\kappa$ plays the role of a scaling parameter related to the mass. At this point we would like to remind the reader that the dynamics of the Hirota model given by the factorized time propagator \eref{propagator} can be interpreted in terms of a two-step Floquet-like protocol -- see \ref{Floq}. 

\section{Integrability}\label{Integrals}

As was shown by Faddeev and Volkov \cite{faddeev,volkov}, the Hirota model described above is integrable. The solution of the functional relation \eref{r_condition} is an \emph{r}-matrix, defined as
\begin{equation}
r(\kappa^2,w)=\sum_{k=(-m+1)/2}^{(m-1)/2}\,\prod_{l=1}^{|k|}\frac{\kappa^2 q^{-l+1}-q^{l-1}}{\kappa^2 q^l-q^{-l}}\, w^k,\label{r_matrix}
\end{equation}
where $w=uv\otimes u v^{-1}$ is the dynamical variable introduced before. The term with index $k=0$ should be thought of as an identity operator acting on $\mathcal{H}^{\otimes 2}$. It can easily be checked that $r(\kappa^2,w)r(\kappa^2,w)^\dagger$ is proportional to an identity if $\kappa\in\mathbbm{R}$, hence the time propagation \eref{propagator} can, in this case, be made unitary. The time propagation, in the case of periodic boundary conditions allows for the two trivial multiplicative conserved charges, namely~\cite{faddeev}
\begin{equation}
I_{\rm even}=\prod_{n=1}^Nw_{2n},\hspace{2cm}I_{\rm odd}=\prod_{n=1}^Nw_{2n-1}.
\end{equation}

Let us now review the algebraic Bethe ansatz construction of the nontrivial conserved charges. The basic ingredient is the Lax operator with legs in both, the \emph{auxiliary space} $\mathcal{V}\cong\mathbbm{C}^2$ equipped with standard, canonically ordered basis $\{\ket{0},\ket{1}\}$, and the copy of the \emph{physical space} $\mathcal{H}_j$ onto which the Weyl variables $u_j,v_j$ act. Taking the well known variant of this Lax operator (see e.g. Refs.~\cite{faddeev,volkov,antonov})
\begin{equation}
L_j(\lambda)=\ket{0}\bra{0}\otimes u_j+\ket{1}\bra{1}\otimes u_j^{-1}+\lambda\ket{0}\bra{1}\otimes v_j-\lambda \ket{1}\bra{0}\otimes v_j^{-1},\qquad 1\le j\le 2N\label{lax}
\end{equation}
and using the intertwining relation
\begin{equation}
L_j(\lambda/\kappa)L_{j-1}(\lambda\kappa)r(\kappa^2,w_j)=r(\kappa^2,w_j)L_{j}(\lambda\kappa)L_{j-1}(\lambda/\kappa),
\end{equation}
one can show that a continuous set of quantities
\begin{equation}
T(\lambda)=\tr_\mathcal{V}\big(L_{2N}(\lambda/\kappa)L_{2N-1}(\lambda \kappa)L_{2N-2}(\lambda/\kappa)L_{2N-3}(\lambda \kappa)...L_2(\lambda/\kappa)L_1(\lambda \kappa)\big)\label{conserved}
\end{equation}
commutes with the time propagator, $\big[U,T(\lambda)\big]=0$, and hence is conserved.\footnote{In the text, $\tr_\mathcal{V}$ denotes the partial trace with respect to the auxiliary space.} Moreover, using the trigonometric $R$-matrix of the $XXZ$ model, one can show the commutativity of these conserved charges, $\big[T(\lambda),T(\mu)\big]=0$. It can easily be proven, that these conserved charges, as well as their derivatives, are either trivial or highly non-local and in particular are not linearly extensive in the system size in the sense of the Hilbert-Schmidt norm. Using the tools that will be described in Section \ref{QL}, one can compute the auxiliary transfer matrix of the Faddeev's charges to get, for even $n\ge 2$
\begin{equation}
\left\| \left[\frac{\rmd^n}{\rmd\lambda^n}T(\lambda)\right]_{\lambda=0} \right\|^2_{\rm HS} \sim N^n.
\end{equation}
More precisely, the dependence on $N$ is polynomial with terms up to the order of the derivative of the transfer operator. Charges that are odd derivatives are all zero. Moreover we have
\begin{equation}
T(0) = I_{\rm odd}\,I_{\rm even} + (I_{\rm odd}\,I_{\rm even})^2,\qquad \|T(0)\|^2_{\rm HS} = 2.
\end{equation}

\section{Quasilocal integrals of motion}\label{QL}

In this section we describe the construction of the quasilocal conservation laws from the Faddeev-Volkov transfer operators. The procedure is somewhat analogous to the one presented in \cite{IMP15} for the case of the isotropic Heisenberg spin chain. 

\subsection{The conjecture and some rigorous arguments}

Let $q=\exp(i\frac{\ell}{m}\pi)$ be a root of unity of an odd order $m$ ($\ell\le m$, $\ell$ even), its square root chosen as $q^{\pm\frac{1}{2}}=\exp(\pm i\frac{\ell}{2m}\pi)$, and let the transfer operator $T(\lambda)$ be given by \eref{conserved}. Additionally, let us denote 
\begin{equation}
\Lambda_s(\lambda)=1+(\kappa^2+\kappa^{-2})\lambda^2+\lambda^4,
\end{equation}
$\kappa$ being the scaling parameter of the Hirota model. The size of the system is taken to be $2N$.
\newline

\noindent\textbf{Conjecture:}
{\it
The conserved charge 
\begin{equation}
X(\lambda)=\frac{1}{\Lambda_s(\lambda)^N}\,T(\lambda q^{\frac{1}{2}})\frac{\rmd}{\rmd\lambda}T(\lambda q^{-\frac{1}{2}})
\label{charge}
\end{equation}	
is a quasilocal operator for $\lambda$ in
\begin{equation}
\mathcal{D}_{q}=\{z\in\mathbbm{C}\setminus\{0\}\mid\arg{z}\in(\pi-\eta\frac{\pi}{2m},\pi+\eta\frac{\pi}{2m})\cup(-\eta\frac{\pi}{2m},\eta\frac{\pi}{2m}),\,\,\eta=\min(\ell, m-\ell)\}.
\end{equation}
}

\subsubsection{Notation and prerequisites} In order to demonstrate the validity of this conjecture, we need to define the structure on the operator space, which allows us to compute Hilbert-Schmidt inner products and norms of operators in a convenient manner. The physical operator space $\End(\mathcal{H})$ can be equipped with an orthonormal basis
\begin{equation}
{\rm e}_{i,j}=u^iv^j,\qquad\langle {\rm e}_{i,j},{\rm e}_{k,l}\rangle=\frac{1}{m}\Tr{\left[({\rm e}_{i,j})^\dagger {\rm e}_{k,l}\right]}=\delta_{ik}\delta_{jl},\quad i,j,k,l\in \mathbbm{Z}_m,\label{operatorbasis}
\end{equation}
where indices differing for the order of the root of unity, $m$, are equivalent due to the cyclicity. Now we can introduce the \emph{auxiliary transfer matrix} $\mathbbm{T}(\lambda_1,\lambda_2,\mu_1,\mu_2)\in\End(\mathcal{V}^{\otimes 4})$ such, that
\begin{equation}
\frac{\Tr{\Big[\big(T(\lambda_1)T(\lambda_2)\big)^\dagger T(\mu_1)T(\mu_2)}\Big]}{\Tr{\mathbbm{1}}}=\Tr{\Big[\mathbbm{T}(\lambda_1,\lambda_2,\mu_1,\mu_2)^N\Big]}\label{HS_1}
\end{equation}
holds for arbitrary complex parameters $\lambda_1,\lambda_2,\mu_1,\mu_2$. The exact definition of the auxiliary transfer matrix is given in \ref{hierarchy}. In short, it can be written in terms of the \emph{double Lax components} $\mathbbm{L}^{[i,j,k,l]}\in\End(\mathcal{V}^{\otimes 2})$ related to the auxiliary components of the ordinary Lax operator \eref{lax}:
\begin{equation}
\mathbbm{T}(\lambda_1,\lambda_2,\mu_1,\mu_2)=\sum_{i,j,k,l}\overline{\mathbbm{L}^{[i,j,k,l]}(\lambda_1,\lambda_2)}\otimes\mathbbm{L}^{[i,j,k,l]}(\mu_1,\mu_2).\label{HS_2}
\end{equation}
Of particular importance is the \emph{leading (zeroth) Lax component} $\mathbbm{L}^{[0,0,0,0]}\equiv\mathbbm{L}_0$ at the special choice of the spectral parameters
\begin{equation}
\mathbbm{L}_0(\lambda q^{\frac{1}{2}},\lambda q^{-\frac{1}{2}})=\left(1+\lambda^4\right)\Big(\ket{01}\bra{01}+\ket{10}\bra{10}\Big)-(\kappa^2+\frac{1}{\kappa^2})\lambda^2\Big(\ket{01}\bra{10}+\ket{10}\bra{01}\Big),
\end{equation}
where $\{\ket{00},\ket{01},\ket{10},\ket{11}\}$ is the canonically ordered basis of $\mathcal{V}^{\otimes 2}$ (for the details see \ref{hierarchy}). It has two nontrivial eigenpairs of which only one is particularly important to this discussion, namely the \emph{singlet eigenpair} $\mathbbm{L}_0(\lambda q^{\frac{1}{2}},\lambda q^{-\frac{1}{2}})\ket{\psi_s}=\Lambda_s(\lambda)\ket{\psi_s}$,
\begin{equation}
\Lambda_s(\lambda)\equiv\Lambda_s(\lambda q^{\frac{1}{2}},\lambda q^{-\frac{1}{2}})=1+(\kappa^2+\frac{1}{\kappa^2})\lambda^2+\lambda^4,\qquad\ket{\psi_s}=\frac{1}{\sqrt{2}}\left(\ket{01}-\ket{10}\right).\label{singlet}
\end{equation}

\subsubsection{The factorizability and the conditions for quasilocality} We are now ready to state two lemmas concerning what is called the \emph{factorizability} of the auxiliary transfer matrix and the singlet eigenpair, as well as the conditions for quasilocality of $X(\lambda)$.
\newline

\noindent\textbf{Lemma 1:}
{\it
The auxiliary transfer matrix satisfies the factorizability condition, namely, $\tau(\lambda,\mu)=\overline{\Lambda_s(\lambda)}\Lambda_s(\mu)$ is its factorized eigenvalue and $\ket{\Psi_s}=\ket{\psi_s}\otimes\ket{\psi_s}$ the corresponding factorized eigenvector, so that:
\eqnarray
\mathbbm{T}(\lambda q^{\frac{1}{2}},\lambda q^{-\frac{1}{2}},\mu q^{\frac{1}{2}},\mu q^{-\frac{1}{2}})\ket{\Psi_s}
=\overline{\mathbbm{L}_0(\lambda q^{\frac{1}{2}},\lambda q^{-\frac{1}{2}})}\ket{\psi_s}\otimes\mathbbm{L}_0(\mu q^{\frac{1}{2}},\mu q^{-\frac{1}{2}})\ket{\psi_s}=\tau(\lambda,\mu)\ket{\Psi_s}.\label{explicitfactorization}
\endeqnarray
}
\bew{(Lemma 1)} Equation \eref{HS_2} implies that the factorizability, as stated in Lemma 1, certainly occurs if
\begin{equation}
\mathbbm{L}^{[i,j,k,l]}(\lambda_1,\lambda_2)\ket{\psi_s}=0,\quad\forall\,\,i+j+k+l>0\label{factor}
\end{equation}
since, in this case, only the leading part $\overline{\mathbbm{L}_0(\lambda q^{\frac{1}{2}},\lambda q^{-\frac{1}{2}})}\otimes\mathbbm{L}_0(\mu q^{\frac{1}{2}},\mu q^{-\frac{1}{2}})$ remains of the whole auxiliary transfer matrix and $\ket{\Psi_s}$ is its eigenstate. An explicit calculation of the double Lax components \eref{two_point}, along with the $m$-independence of \eref{stag_comps} implies that the condition \eref{factor} is $m$-independent. Putting $\lambda_1=z q^\alpha$, $\lambda_2=z q^{\beta}$ with $z\in\mathbbm{C}$, the factorization condition \eref{factor} becomes simply $q^{1-\alpha+\beta}=1$ and gives the final result $\lambda_1=\lambda q^{\frac{1}{2}}$, $\lambda_2=\lambda q^{-\frac{1}{2}}$, used in \eref{explicitfactorization}. The factorization thus occurs at the relatively shifted spectral parameters, similarly as in the case of the isotropic Heisenberg model \cite{IMP15}.\hfill$\square$
\endbew

\noindent\textbf{Lemma 2:} 
{\it
Let $\Lambda_s(\lambda)$ and $\tau(\lambda)\equiv\tau(\lambda,\lambda)$ be the isolated leading eigenvalues of $\mathbbm{L}_0(\lambda q^{\frac{1}{2}},\lambda q^{-\frac{1}{2}})$ and $\mathbbm{T}(\lambda q^{\frac{1}{2}},\lambda q^{-\frac{1}{2}},\lambda q^{\frac{1}{2}},\lambda q^{-\frac{1}{2}})$ respectively and let both of these operators be diagonalizable. Then, in the thermodynamic limit, $X(\lambda)$ given by \eref{charge} scales linearly in the system size, namely $\|X(\lambda)\|^2_{\rm HS}\propto N$.
}
\newline

\noindent\textbf{Remark:} By the \emph{isolated leading eigenvalue} we mean an eigenvalue, which is maximal in absolute value and is separated from the rest of the spectrum by a gap.
\bew{(Lemma 2)} Using the definition of the Hilbert-Schmidt inner product \eref{def_hs} we have 
\eqnarray
\fl\|X(\lambda)\|^2_{\rm HS}=\frac{1}{\tau(\lambda)^N}\Big[\frac{\partial^2}{\partial \overline{x}\partial y}\tr{\big[\mathbbm{T}(\lambda q^{\frac{1}{2}}, x,\lambda q^{\frac{1}{2}}, y)^N\big]}
-\frac{\partial}{\partial \overline{x}}\overline{\tr{\big[\mathbbm{L}_0(\lambda q^{\frac{1}{2}},x)^N\big]}}\frac{\partial}{\partial y}\tr{\big[\mathbbm{L}_0(\lambda q^{\frac{1}{2}},y)^N\big]}\Big]_{x,y=\lambda q^{-\frac{1}{2}}}.
\endeqnarray
In the second term we have used the fact that only the identity component of $T(\lambda q^\frac{1}{2})T(\lambda q^{-\frac{1}{2}})$ has a non-vanishing trace. By the assumption, there is a square matrix $S(\lambda)$ such that $S(\lambda)^{-1}\mathbbm{T}(\lambda q^{\frac{1}{2}},\lambda q^{-\frac{1}{2}},\lambda q^{\frac{1}{2}},\lambda q^{-\frac{1}{2}})S(\lambda)$ is a diagonal matrix. The trace of an arbitrary operator $\mathbbm{A}\in\End(\mathcal{V}^{\otimes 4})$ can be rewritten as
\begin{equation}
\tr{\mathbbm{A}}=\sum_{\boldsymbol n}\braket{\boldsymbol n|\mathbbm{A}|\boldsymbol n}=\braket{\widetilde{\Psi}_s|\mathbbm{A}|\Psi_s}+\sum_{\boldsymbol n\ne \boldsymbol m}\braket{\boldsymbol n|S(\lambda)^{-1}\mathbbm{A}\,S(\lambda)|\boldsymbol n},\label{trace_redefinition}
\end{equation}
where, for some element $\ket{\boldsymbol m}$ of the orthonormal basis $\{\ket{\boldsymbol n}=\ket{n_1,n_2,n_3,n_4}\}$ of $\mathcal{V}^{\otimes 4}$, the two vectors
\begin{equation}
\bra{\widetilde{\Psi}_s}\equiv\bra{\boldsymbol m}S(\lambda)^{-1},\qquad\ket{\Psi_s}\equiv S(\lambda)\ket{\boldsymbol m}\label{left_right}
\end{equation}
are the left and the right eigenvectors of the auxiliary transfer matrix corresponding to the eigenvalue $\tau(\lambda)$. Note that the second term of \eref{trace_redefinition} contains only left and right eigenvectors of the auxiliary transfer matrix, corresponding to the non-leading eigenvalues. Using \eref{trace_redefinition} and taking into account the assumption that $\tau(\lambda)$ and $\Lambda_s(\lambda)$ are the leading eigenvalues we now get\footnote{We should remark that due to representation \eref{HS_2} the auxiliary transfer matrix is antiholomorphic in the first two variables, hence the partial derivative on $\overline{x}$ is nontrivial.}
\eqnarray
\fl\|X(\lambda)\|^2_{\rm HS}=
\!\frac{N}{\tau(\lambda)}
\!\Bigg[\!\bra{\widetilde{\Psi}_s}\frac{\partial^2}{\partial \overline{x}\partial y}\mathbbm{T}(\lambda q^{\frac{1}{2}}, x,\lambda q^{\frac{1}{2}},y)\ket{\Psi_s}
-\frac{\bra{\widetilde{\Psi}_s}\frac{\partial}{\partial \overline{x}}\mathbbm{T}(\lambda q^{\frac{1}{2}}, x,\lambda q^{\frac{1}{2}}, \lambda q^{-\frac{1}{2}})\frac{\partial}{\partial y}\mathbbm{T}(\lambda q^{\frac{1}{2}}, \lambda q^{-\frac{1}{2}},\lambda q^{\frac{1}{2}},y)\ket{\Psi_s}}{\tau(\lambda)}\!\cr\cr\cr
\fl+\,N\bigg(\frac{\bra{\widetilde{\Psi}_s}\frac{\partial}{\partial \overline{x}}\mathbbm{T}(\lambda q^{\frac{1}{2}}, x,\lambda q^{\frac{1}{2}}, \lambda q^{-\frac{1}{2}})\frac{\partial}{\partial y}\mathbbm{T}(\lambda q^{\frac{1}{2}}, \lambda q^{-\frac{1}{2}},\lambda q^{\frac{1}{2}},y)\ket{\Psi_s}}{\tau(\lambda)}
-\frac{\partial}{\partial \overline{x}}\overline{\Lambda_s(\lambda q^{\frac{1}{2}},x)}\frac{\partial}{\partial y}\Lambda_s(\lambda q^{\frac{1}{2}},y)\bigg)\!\Bigg]_{x,y=\lambda q^{-\frac{1}{2}}}\cr\cr\cr
\fl+\mathcal{O}(e^{-\gamma N}).\label{HSnorm_lema}
\endeqnarray
Here $\gamma>0$ is the logarithm of the absolute value of the ratio between $\tau(\lambda)$ and the second-to-leading eigenvalue. $\Lambda_s(\lambda q^{\frac{1}{2}},x)$ can be computed from \eref{sin} since $\mathbbm{L}_0(\lambda q^{\frac{1}{2}},x)\ket{\psi_s}=\Lambda_s(\lambda q^{\frac{1}{2}},x)\ket{\psi_s}$, as $\ket{\psi_s}$ is parameter-independent. Equation \eref{HS_2} along with $\mathbbm{L}^{[i,j,k,l]}(\lambda q^{\frac{1}{2}},\lambda q^{-\frac{1}{2}})\ket{\psi_s}=0,\quad\forall\,\,i+j+k+l>0$ (see the proof of Lemma 1) now imply 
\begin{equation}
\frac{\bra{\widetilde{\Psi}_s}\frac{\partial}{\partial \overline{x}}\mathbbm{T}(\lambda q^{\frac{1}{2}}, x,\lambda q^{\frac{1}{2}}, \lambda q^{-\frac{1}{2}})\frac{\partial}{\partial y}\mathbbm{T}(\lambda q^{\frac{1}{2}}, \lambda q^{-\frac{1}{2}},\lambda q^{\frac{1}{2}},y)\ket{\Psi_s}}{\tau(\lambda)}=\frac{\partial}{\partial \overline{x}}\overline{\Lambda_s(\lambda q^{\frac{1}{2}},x)}\frac{\partial}{\partial y}\Lambda_s(\lambda q^{\frac{1}{2}},y)\label{cancellation}
\end{equation}
and thus \eref{HSnorm_lema} becomes linear in $N$ with exponentially decaying correction,
\begin{equation}
\fl\|X(\lambda)\|^2_{\rm HS}=
\frac{N}{\tau(\lambda)}\Bigg[\!\bra{\widetilde{\Psi}_s}\frac{\partial^2}{\partial \overline{x}\partial y}\mathbbm{T}(\lambda q^{\frac{1}{2}}, x,\lambda q^{\frac{1}{2}},y)\ket{\Psi_s}-
\frac{\partial}{\partial \overline{x}}\overline{\Lambda_s(\lambda q^{\frac{1}{2}},x)}\frac{\partial}{\partial y}\Lambda_s(\lambda q^{\frac{1}{2}},y)\Bigg]_{x,y=\lambda q^{-\frac{1}{2}}}\!\!+\mathcal{O}(e^{-\gamma N}).
\end{equation}
For large system sizes the correction vanishes and only the linear dependence remains.
\hfill$\square$
\endbew

\subsubsection{Validity of the conjecture} Let us finally deal with the validity of the conjecture. We need to consider the domain of the spectral parameter $\lambda$ in which: (1)~$\Lambda_s(\lambda)$ is the isolated leading eigenvalue of $\mathbbm{L}_0(\lambda q^{\frac{1}{2}},\lambda q^{-\frac{1}{2}})$, (2)~$\tau(\lambda)=|\Lambda_s(\lambda)|^2$ is the isolated leading eigenvalue of $\mathbbm{T}(\lambda q^{\frac{1}{2}},\lambda q^{-\frac{1}{2}},\lambda q^{\frac{1}{2}},\lambda q^{-\frac{1}{2}})$. Lemmas 1 and 2 then imply the linear extensivity, i.e., quasilocality of the conservation laws.

Recall that the leading Lax component has two nontrivial eigenvalues \eref{sin}, \eref{tri}. The first one, $\Lambda_s(\lambda)$ as given by \eref{singlet}, is the leading one in the absolute value when $\lambda$ is in
\begin{equation}
\mathcal{D}_>=\{z\in\mathbbm{C}\setminus\{0\}\mid \arg z\in(-\frac{\pi}{4},\frac{\pi}{4})\cup(\pi-\frac{\pi}{4},\pi+\frac{\pi}{4})\}.
\end{equation}
The auxiliary transfer matrix $\mathbbm{T}(\lambda q^{\frac{1}{2}},\lambda q^{-\frac{1}{2}},\lambda q^{\frac{1}{2}},\lambda q^{-\frac{1}{2}})$ can be calculated explicitly and decomposed as 
\begin{equation}
\mathbbm{T}(\lambda q^{\frac{1}{2}},\lambda q^{-\frac{1}{2}},\lambda q^{\frac{1}{2}},\lambda q^{-\frac{1}{2}})=\mathbbm{T}^{\rm (r)}(\lambda,\lambda)\oplus 0.
\end{equation} 
The null space is spanned by ten vectors $\ket{n_1,n_2,n_3,n_4} \in \mathcal{V}^{\otimes 4}$, with $n_1+n_2 \neq n_3+n_4$, for which
$ \mathbbm{T}(\lambda q^{\frac{1}{2}},\lambda q^{-\frac{1}{2}},\lambda q^{\frac{1}{2}},\lambda q^{-\frac{1}{2}}) \ket{n_1,n_2,n_3,n_4} =0$ and
$ \bra{n_1,n_2,n_3,n_4} \mathbbm{T}(\lambda q^{\frac{1}{2}},\lambda q^{-\frac{1}{2}},\lambda q^{\frac{1}{2}},\lambda q^{-\frac{1}{2}}) =0$ holds.
 
\begin{figure}[h]
\includegraphics[width=450pt]{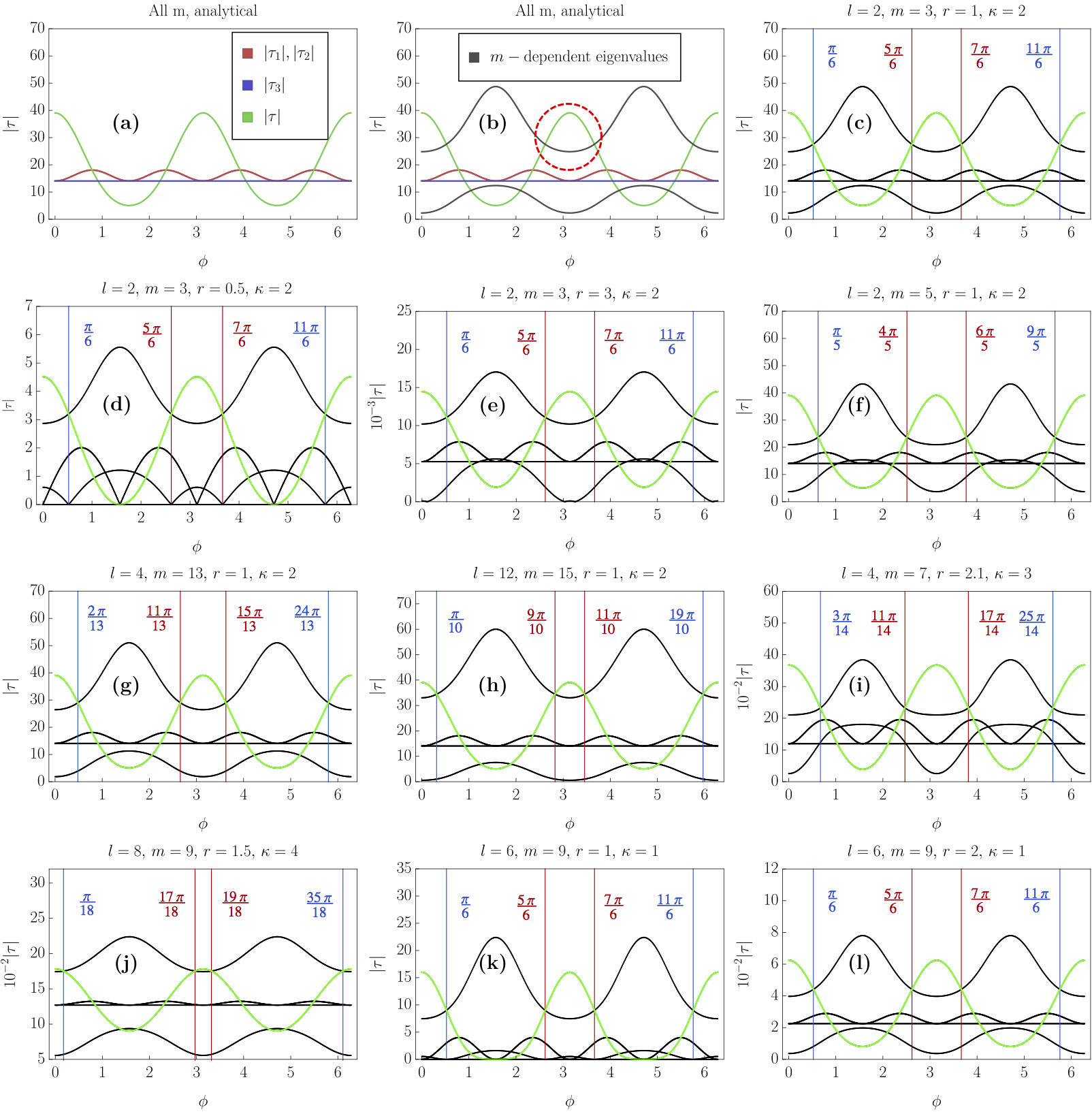}
\caption{$\phi$-dependence of the eigenvalues of the reduced auxiliary transfer matrix $\mathbbm{T}^{\rm (r)}(r e^{i\phi},r e^{i\phi})$ for $r=0.5,\,1,\,1.5,\,2,\,2.1,\,3$, $\kappa=1,\,2,\,3,\,4$. For $m=3$ they can be computed analytically -- see diagram {\rm \textbf{(b)}}. Four of these eigenvalues are $m$-independent and are shown on the diagram {\rm \textbf{(a)}}. The important part of the diagram, where $\tau(r e^{i\phi})$ (green colour) becomes the leading eigenvalue, is marked with a red dashed circle on the diagram {\rm \textbf{(b)}} and explicitly labeled with the vertical red and blue lines on the rest of the diagrams. The diagrams {\rm \textbf{(c)}}-{\rm \textbf{(l)}} hint at the domains of the complex spectral parameter $\lambda$, for which $\tau(\lambda)$ is the leading eigenvalue of the auxiliary transfer matrix. Additionally note, that the two non-leading eigenvalues (panel {\rm \textbf{(a)}}, dark red) are degenerate in the absolute value. The diagrams are consistent with the statement that the domains of quasilocality in the complex plane are wedges $\arg{\lambda}\in(\pi-\eta\frac{\pi}{2m},\pi+\eta\frac{\pi}{2m})\cup(-\eta\frac{\pi}{2m},\eta\frac{\pi}{2m})$ with $\eta=\min(\ell, m-\ell)$. The point $\lambda=0$ should be excluded due to the degeneracy of the reduced auxiliary transfer matrix, resulting in a high non-locality of the conservation laws. This kind of numerics seems to confirm the domain $\mathcal{D}_{q}$ for all $\kappa$ and $q$.}
\label{fig:numerics}
\end{figure}

\begin{figure}[h]
\begin{minipage}{0.45\linewidth}
\includegraphics[width=190pt]{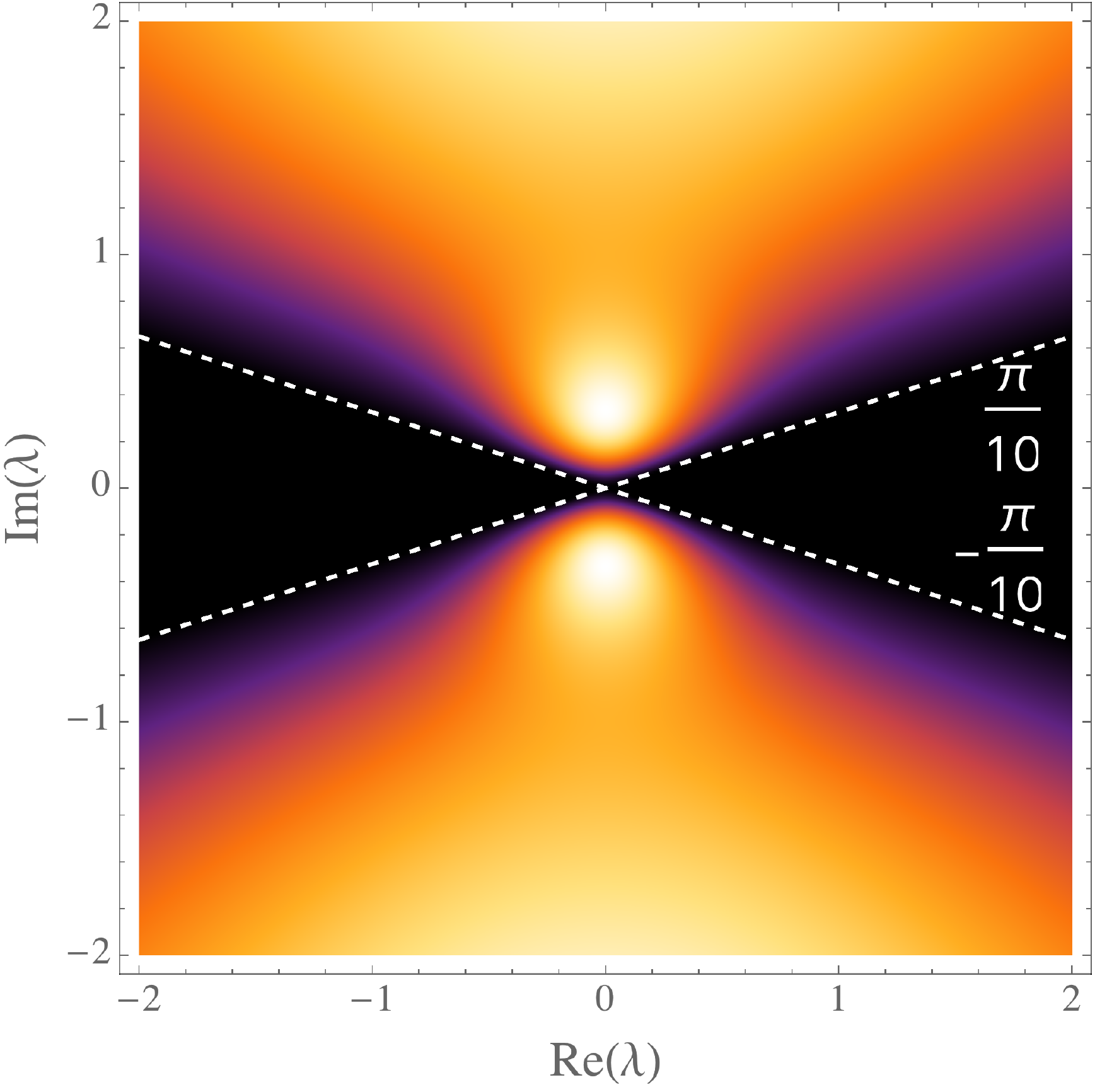}
\end{minipage}
\begin{minipage}{0.45\linewidth}
\includegraphics[width=190pt]{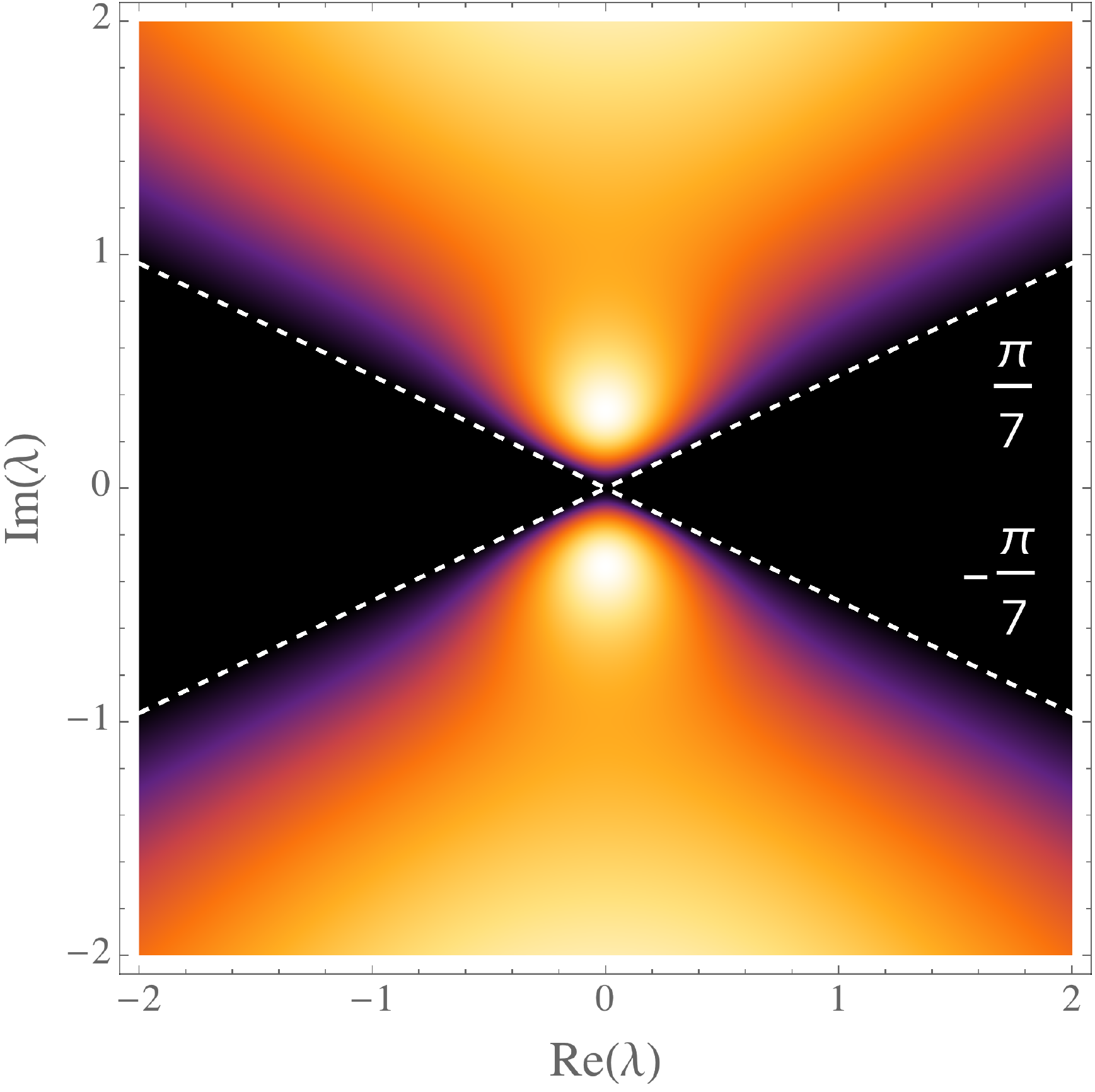}
\end{minipage}
\caption{Wedges in the complex plane, where $\tau(\lambda)$ is the leading eigenvalue of the auxiliary transfer matrix. Left figure shows the case $q=e^{i4\pi/5}$, while the right one $q=e^{i2\pi/7}$. In both cases the scaling parameter has the value $\kappa=3$. Plotted is the appropriately colour-coded function  $1-\frac{\tau(\lambda)}{\rho(\mathbbm{T}^{\rm (r)}(\lambda,\lambda))}$, black for the value 0, where $\rho(\mathbbm{T}^{\rm (r)}(\lambda,\lambda))$ denotes the spectral radius of the auxiliary transfer matrix. Analogous situation occurs in the case of the Heisenberg model \cite{P11,IMP15}, where the wedge in the complex plane becomes a strip, since in that case the spectral parameter is represented as $\lambda=e^{i\varphi}$, $\varphi\in\mathbbm{C}$.}\label{fig:cones}
\end{figure}

The explicit form of the nontrivial $6\times 6$ \emph{reduced auxiliary transfer matrix} $\mathbbm{T}^{\rm (r)}(\lambda,\lambda)$ is given in \ref{Matr}. It is diagonalizable and its spectrum contains four $q$-independent eigenvalues,
\eqnarray
&\tau_1(\lambda)=\big(\overline{\lambda}^2-\kappa^2\big) \big(\lambda^2+\kappa^2\big) \big(\overline{\lambda}^2 -\frac{1}{\kappa^2}\big) \big(\lambda^2+\frac{1}{\kappa^2}\big),\label{deg1}\\
&\tau_2(\lambda)=\big(\overline{\lambda} ^2+\kappa^2\big) \big(\lambda^2-\kappa^2\big) \big(\overline{\lambda} ^2+\frac{1}{\kappa^2}\big) \big(\lambda^2 -\frac{1}{\kappa^2}\big),\label{deg2}\\
&\tau_3(\lambda)=1-\big(\kappa^4+\frac{1}{\kappa^4}\big)|\lambda|^4+|\lambda|^8,\\
&\tau(\lambda)=|\big(\lambda ^2+\kappa^2\big) \big(\lambda ^2+\frac{1}{\kappa^2}\big)|^2=|\Lambda_s(\lambda)|^2,
\endeqnarray
as well as additional two $q$-dependent eigenvalues, which, at present, we are unable to write down explicitly for an arbitrary root of unity $q$. 
At this point we have to use numerical analysis in order to demonstrate the conjecture. For some $\lambda$, one of these two $q$-dependent eigenvalues exceeds $\tau(\lambda)$ in the absolute value, while all the other eigenvalues are smaller. This restricts the spectral parameter $\lambda$, for which $\tau(\lambda)$ is the leading eigenvalue, onto the domain $\mathcal{D}_{q}$ given by the conjecture. The domain $\mathcal{D}_{q}$ is deduced using the exact numerical diagonalization of the reduced auxiliary transfer matrix $\mathbbm{T}^{\rm (r)}(\lambda,\lambda)$ (see figures~\ref{fig:numerics},~\ref{fig:cones} and their captions for more details). Note that for each root of unity $q$ we have $\mathcal{D}_q\subset \mathcal{D}_>$. For $\lambda\in\mathcal{D}_q$, $\Lambda_s(\lambda)$ is thus automatically the isolated leading eigenvalue of the leading Lax component $\mathbbm{L}_0(\lambda q^{\frac{1}{2}},\lambda q^{-\frac{1}{2}})$.

We should remark that there is another nontrivial eigenpair of the leading double Lax component -- the triplet eigenpair $\Lambda_t(\lambda)$ and $\ket{\psi_t}$, \eref{tri}.
Similar analysis as above shows that although the reduced auxiliary transfer matrix is different in this case, its spectrum is identical. It turns out that the factorization of the triplet eigenpair gives exactly the same conservation laws, $X(\lambda)$, therefore it will not be considered separately. With figures~\ref{fig:numerics},~\ref{fig:cones} we conclude the discussion of the conjecture.
\clearpage
\clearpage

\subsection{Symmetries of the conserved charges and the auxiliary transfer matrix}

Here we note several interesting observations. Firstly, the auxiliary transfer matrix (see \ref{Matr}) is invariant under the exchange $\kappa\mapsto\frac{1}{\kappa}$. Secondly, symbolic manipulation shows that the auxiliary transfer matrices with different $\kappa$ commute, i.e.
\begin{equation}
\big[\mathbbm{T}^{\rm (r)}(\lambda,\mu;\kappa),\mathbbm{T}^{\rm (r)}(\lambda,\mu;\kappa')\big]=0.
\end{equation}
Moreover, the auxiliary transfer matrix possesses a nice symmetry $[{\rm\mathbf P},\mathbbm{T}^{\rm (r)}(\lambda,\mu)]=0$, where ${\rm\mathbf P}$ is a permutation matrix with ones on the anti-diagonal, i.e.
\begin{equation}
{\rm\mathbf P}=\pmatrix{
0 & 0 & 0 & 0 & 0 & 1\cr
0 & 0 & 0 & 0 & 1 & 0\cr
0 & 0 & 0 & 1 & 0 & 0\cr
0 & 0 & 1 & 0 & 0 & 0\cr
0 & 1 & 0 & 0 & 0 & 0\cr
1 & 0 & 0 & 0 & 0 & 0},
\qquad
{\rm\mathbf P}^2=\mathbbm{1}.\label{parity}
\end{equation}
This invariance of the reduced auxiliary transfer matrix is connected to the parity symmetry $\mathcal{P}$ of the conserved quantities, defined on the physical space as
\begin{equation}
u\mapsto u^{-1},\qquad v\mapsto v^{-1}.
\end{equation}
One notes that conjugation by $P=-\sigma^y\otimes\sigma^y$ (we use the standard notation for the Pauli matrices) is the corresponding transformation on the auxiliary space in a sense
\begin{equation}
P\mathbbm{L}^{[i,j,k,l]}(\lambda_1,\lambda_2)P\otimes \mathcal{P}\left({\rm e}_{i,j}\otimes {\rm e}_{k,l}\right)=\mathbbm{L}^{[\mn i,\mn j,\mn k,\mn l]}(\lambda_1,\lambda_2)\otimes {\rm e}_{\mn i,\mn j}\otimes {\rm e}_{\mn k,\mn l},
\end{equation}
where indices should be taken modulo $m$. Since the full double Lax operator contains both $\mathbbm{L}^{[i,j,k,l]}(\lambda_1,\lambda_2)$, as well as $\mathbbm{L}^{[\mn i,\mn j,\mn k,\mn l]}(\lambda_1,\lambda_2)$, we have
\begin{equation}
P \mathbbm{L}(\lambda_1,\lambda_2) P = {\cal P}\left(  \mathbbm{L}(\lambda_1,\lambda_2)\right)
\end{equation}
and hence the operators $T(\lambda_1)T(\lambda_2)$ are parity symmetric. The parity transformation is parameter independent, thus the logarithmic derivatives $X(\lambda)$ also possess the parity symmetry
\begin{equation}
X(\lambda) = {\cal P}\left(X(\lambda)\right).
\end{equation}
The parity transformation can now be given sense in the context of the auxiliary transfer matrix. There one needs $P\otimes P$ to act on the double auxiliary space. Restriction of this operator onto the $6-$dimensional subspace, where $\mathbbm{T}$ is nonzero, gives ${\rm\mathbf P}$ as defined in \eref{parity}.

\section{The Hilbert-Schmidt kernel and the matrix product representation}\label{Kernel}

\subsection{The Hilbert-Schmidt kernel explicitly}

The results of the previous section now imply the existence of a kernel $\mathcal{K}(\lambda,\mu)$, such that the following equation holds
\begin{equation}
\langle X(\lambda),X(\mu)\rangle=N\,\mathcal{K}(\lambda,\mu)+\mathcal{O}(e^{-\gamma N}),\qquad \gamma>0\label{linearextensivity}
\end{equation}
in the wedge of quasilocality $\mathcal{D}_q$. For spectral parameters $\lambda,\mu\in\mathcal{D}_q$ we can calculate the \emph{Hilbert-Schmidt kernel} similarly as in the proof of Lemma 2, to get
\begin{equation}
\fl\mathcal{K}(\lambda,\mu)=\frac{1}{\tau(\lambda,\mu)}\Bigg[\!\bra{\widetilde{\Psi}_s}\frac{\partial^2}{\partial \overline{x}\partial y}\mathbbm{T}(\lambda q^{\frac{1}{2}}, x,\mu q^{\frac{1}{2}},y)\ket{\Psi_s}-\frac{\partial}{\partial \overline{x}}\overline{\Lambda_s(\lambda q^{\frac{1}{2}},x)}\frac{\partial}{\partial y}\Lambda_s(\mu q^{\frac{1}{2}},y)\Bigg]_{x=\lambda q^{-\frac{1}{2}},y=\mu q^{-\frac{1}{2}}}.\label{hsk1}
\end{equation}
Note, that this now holds at two spectral parameters, $\lambda$ and $\mu$. In deriving \eref{hsk1} we have taken into account the fact that in the wedge of quasilocality, $\mathcal{D}_q$, $\tau(\lambda,\mu)$ is the leading eigenvalue of $\mathbbm{T}^{\rm (r)}(\lambda,\mu)$. This follows from $\tau(\lambda)$ being the leading eigenvalue of $\mathbbm{T}^{\rm (r)}(\lambda,\lambda)$ in the same wedge, a result of the previous section. Indeed, if $\tau(\lambda,\mu)$ were not of the maximal absolute value, an $\mathcal{O}(N^2)$ contribution would be present in \eref{hsk1}, similarly as in \eref{HSnorm_lema}, but due to the other eigenvalues. Because the latter are not factorizable, this $\mathcal{O}(N^2)$ terms would not cancel as in the case of the factorized leading eigenvalue -- see \eref{cancellation}. However, that would contradict the Cauchy-Schwarz inequality, since $X(\lambda)$ scale at most linearly in $N$, as was argued in the previous section. 

\begin{figure}[h]
\includegraphics[width=470pt]{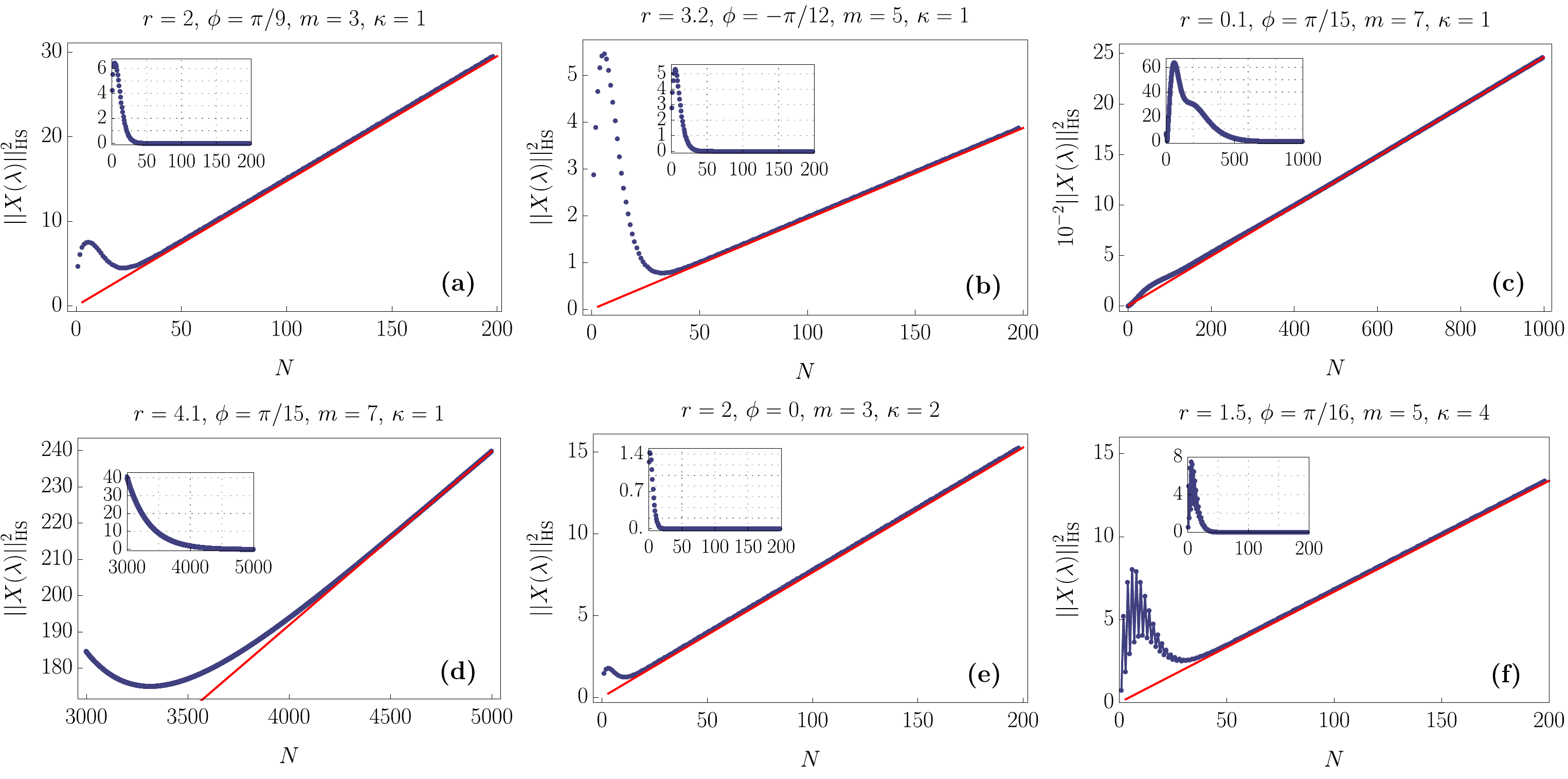}
\caption{The Hilbert-Schmidt norm of $X(\lambda)$ using the Hilbert-Schmidt kernel \eref{HSKernel} (red line) versus the full Hilbert-Schmidt norm of the charge $X(\lambda)$ (dark blue) for $\lambda=\mu=r e^{i\phi}$. From the top left to the bottom right panels, the roots of unity are $e^{i 2\pi/3}$, $e^{i 2\pi/5}$, $e^{i 2\pi/7}$, $e^{i 6\pi/7}$, $e^{i 2\pi/3}$ and again $e^{i 2\pi/5}$. The inset diagrams represent the absolute difference between full (numerical) Hilbert-Schmidt norm and the one computed using our (analytic) expression for the Hilbert-Schmidt kernel. The difference tends to zero in the thermodynamic limit.}
\label{fig:check}
\end{figure}

After a straightforward calculation of the objects involved in \eref{hsk1} we obtain
\begin{equation}
\fl\mathcal{K}(\lambda,\mu)=\frac{\left(2-q^2-\frac{1}{q^2}\right) \overline{\lambda}  \mu  \left(\overline{\lambda} ^2+\mu ^2\right) \left(\left(\kappa^2+\frac{1}{\kappa^2}\right)\left(\overline{\lambda}^2\mu^2+1\right)+2 \overline{\lambda} ^2+2 \mu ^2\right)}{2 \left(\overline{\lambda}^4+\left(\kappa^2+\frac{1}{\kappa^2}\right) \overline{\lambda} ^2+1\right) \left(\mu ^4+\left(\kappa^2+\frac{1}{\kappa^2}\right) \mu ^2+1\right) \left(\overline{\lambda} ^4+\mu ^4-\overline{\lambda} ^2 \mu ^2 \left(q^2+\frac{1}{q^2}\right)\right)}\label{HSKernel}
\end{equation}
for the explicit form of the Hilbert-Schmidt kernel. Looking back at figure~\ref{fig:numerics}, one notes the degeneracy of the two non-leading eigenvalues of the auxiliary transfer matrix at $\lambda=\mu$. It turns out, that this degeneracy poses problems only for $\operatorname{Im}\lambda= 0$ since in this case the matrix $S(\lambda)$, for which $S(\lambda)^{-1}\mathbbm{T}^{\rm (r)}(\lambda,\lambda)S(\lambda)$ is diagonal, becomes singular.\footnote{We use the same symbol, $S(\lambda)$, for both the matrix that diagonalizes $\mathbbm{T}(\lambda q^{\frac{1}{2}}, \lambda q^{-\frac{1}{2}},\lambda q^{\frac{1}{2}},\lambda q^{-\frac{1}{2}})=\mathbbm{T}^{\rm (r)}(\lambda,\lambda)\oplus 0$ as well as the one that diagonalizes $\mathbbm{T}^{\rm (r)}(\lambda,\lambda)$.} However, the result of the numerical check indicates that the explicit form of the Hilbert-Schmidt kernel holds even at $\lambda=\mu\in\mathbbm{R}$. An example is given by diagram $({\rm e})$ of figure~\ref{fig:check}.

\subsection{Matrix product form of quasilocal conserved charges}

Following \cite{IMP15}, we can write down the matrix product ansatz for the conserved quantities $X(\lambda)$ in the thermodynamic limit. Let us consider the Hilbert-Schmidt projection of the conserved charge onto the local basis operator 
\begin{equation}
{\rm\mathbf{e}}_{[\{\textbf{k}_\alpha\}_{\alpha=1}^r]}\equiv{\rm\mathbf{e}}_{[\{(i_\alpha,j_\alpha,k_\alpha,l_\alpha)\}_{\alpha=1}^r]}\equiv\left(\bigotimes_{\alpha=1}^r {\rm e}_{i_\alpha,j_\alpha}\otimes {\rm e}_{k_\alpha,l_\alpha}\right)\otimes \mathbbm{1}^{\otimes (2N-2r)}.
\end{equation}
Since $\Lambda_s(\lambda)$ is the leading eigenvalue of the zeroth Lax component, in the thermodynamic limit, its action amounts to a projector
\begin{equation}
\lim_{n\to\infty}\left(\frac{\mathbbm{L}_0(\lambda q^{\frac{1}{2}},\lambda q^{-\frac{1}{2}})}{\Lambda_s(\lambda)}\right)^n=\ket{\psi_s}\bra{\psi_s}.
\end{equation}
Since all the other Lax components destroy the singlet state, the nontrivial action of the local density of the charge $X(\lambda)$ should start with the derivative of the Lax operator on the right side (similarly as in the case of the Heisenberg chain \cite{IMP15}). In the thermodynamic limit we thus have
\eqnarray
\fl\langle{\rm\mathbf{e}}_{[\{(i_\alpha,j_\alpha,k_\alpha,l_\alpha)\}_{\alpha=1}^r]},X(\lambda)\rangle 
=\frac{\bra{\psi_s}\prod_{\alpha=r}^{2}\mathbbm{L}^{[i_\alpha,j_\alpha,k_\alpha,l_\alpha]}(\lambda q^{\frac{1}{2}},\lambda q^{-\frac{1}{2}})\,\big[\partial_\mu\mathbbm{L}^{[i_1,j_1,k_1,l_1]}(\lambda q^{\frac{1}{2}},\mu q^{-\frac{1}{2}})\big]_{\mu=\lambda}\ket{\psi_s}}{\Lambda_s(\lambda)^r},\\[1em]
\fl X(\lambda)= \sum_{r=2}^N \sum_{\{\textbf{k}_\alpha\}\in\mathbbm{Z}_m^{4r}}\langle {\rm\mathbf{e}}_{[\{\textbf{k}_\alpha\}_{\alpha=1}^r]},X(\lambda)\rangle \sum_{j=0}^{N-1}\hat{\cal S}^{j}\big({\rm\mathbf{e}}_{[\{\textbf{k}_\alpha\}_{\alpha=1}^r]}\big),
\endeqnarray
where $\hat{\cal S}$ is a periodic shift automorphism for two physical sites (due to the staggering of the transfer operator -- see \eref{conserved}), its action being
\begin{equation}
\hat{\cal S}(\mathbbm{1}^{\otimes j}\otimes A\otimes\mathbbm{1}^{\otimes 2N-n-j})=\mathbbm{1}^{\otimes j+2}\otimes A\otimes\mathbbm{1}^{\otimes 2N-n-j-2},\qquad A\in\End(\mathcal{H}^{\otimes n}).
\end{equation}
An explicit computation of the actions of the Lax components onto the singlet state results in the conclusion that the traceless part of the conserved charge $X(\lambda)$ contains only terms acting nontrivially on at least four consecutive physical lattice sites, hence the sum over $r$ starts with $r=2$. \verb|Mathematica| code for constructing the matrix product representation of the quasilocal charges $X(\lambda)$ is available on the web \cite{hirota_nb}.

\section{Conclusion}

In this paper we have established a procedure to construct quasilocal integrals of motion for the quantum Hirota model which can, in short, be described as a Floquet driven chain of interacting finite-dimensional quantum systems. In an appropriate scaling limit the model also describes the quantum sine-Gordon field theory in $1+1$ dimensions. We note that an alternative procedure to construct quasilocal charges in the integrable field theories with non-diagonal scattering, which builds on the discrete light cone approach with fermionic (spin-$1/2$) variables of Destri and De Vega \cite{destri,destri2,takacs}, has been suggested in Ref.~\cite{vernier}.

The quasilocal integrals of motion described in this paper stem from Faddeev-Volkov conservation laws~\cite{faddeev} which are built using the standard procedure of the algebraic Bethe ansatz. In showing that the quasilocality holds, we have followed the procedure put forward in \cite{IMP15}: first we have established the factorizability of the leading eigenpair of the auxiliary transfer matrix, due to which only terms proportional to the system size remain in the Hilbert-Schmidt norm of the conserved quantities. Then we have identified the regions of the spectral parameter for which the quasilocality of these conservation laws holds. The identification of these regions was based on the leadingness of the factorized eigenvalue.
We have seen, that the quasilocality arises when the spectral parameter falls into a wedge in the complex plane, the opening angle of which is determined only by the root-of-unity deformation (quantization) parameter $q$.

Our conservation laws are parity-invariant, as follows from the intrinsic symmetry properties of the Lax operators. Potentially they can be used to define the Generalized Gibbs ensembles in the quantum quench problems 
(see e.g. \cite{spyros} for a discussion of a quench problem for a quantum field theory) and to establish bounds on the dynamical susceptibilities based on the Mazur inequality (building on \cite{P11}).  

\section*{Acknowledgements}

LZ thanks M. Medenjak for the discussions related to the quasilocality in the isotropic Heisenberg model. We thank E. Ilievski and G. Takacs for useful comments on the manuscript. The work has been supported by ERC grant OMNES and Slovenian Research Agency grant N1-0025 and programme P1-0044.

\appendix

\section{Floquet picture of the dynamics}\label{Floq}

Here we briefly describe the Floquet picture of the dynamics. Since the propagator \eref{propagator} consists of two factorized parts and the factors of each part commute among themselves, because they depend either only on the odd-numbered or only on the even-numbered dynamical variables, we can define two effective hamiltonians in the following way:
\eqnarray
&U_{\rm even}=\exp(-i H_{\rm even}),  \hspace{1cm} H_{\rm even}=i\sum_{n=1}^N \log r(\kappa^2,w_{2n}),\\
&U_{\rm odd}=\exp(-i H_{\rm odd}), \hspace{1cm} H_{\rm odd}=i\sum_{n=1}^N \log r(\kappa^2,w_{2n-1}),
\endeqnarray
so that the whole time-propagation is generated by a periodic time-dependent hamiltonian $H(t+1)=H(t)$, defined as $H(0\le t < 1/2) = 2 H_{\rm even}$, $H(1/2 \le t < 1) = 2 H_{\rm odd}$.
If $\kappa$ is real we can normalize $r$-matrices to become unitary. Then the effective hamiltonians get an additive complex constant and become hermitian. Without going into the details let us, just as an example, construct these hamiltonians explicitly for the case $q^3=1$ that is $q=\exp(i\frac{2}{3}\pi)$. The $r$-matrix takes a simple form
\begin{equation}
r(\kappa^2,w)=\one\otimes\one+\frac{\kappa^2-1}{\kappa^2 q-q^{-1}}\left(w+w^{-1}\right)=\one\otimes\one+\alpha(\kappa^2)\,W,\qquad W=w+w^{-1}
\end{equation}
and the local hamiltonian density can be expanded into a logarithmic series.
Since $w^3=1$ the following holds for the powers of $W$:
\eqnarray
W^n=A(n)+B(n) W,
\endeqnarray
where coefficients $A(n)$ and $B(n)$ satisfy recursive relations of the form
\eqnarray
A(n+1)=2B(n),\\[.5em]
B(n+1)=A(n)+B(n),
\endeqnarray
with the initial conditions $A(1)=0, A(2)=2, B(1)=B(2)=1$. These relations can be solved to get the final result
\begin{equation}
\log r(\kappa^2,w)=\frac{1}{3}\log\Big[\big(1+2\alpha(\kappa^2)\big)\big(1-\alpha(\kappa^2)\big)^2\Big]\one\otimes\one+\frac{1}{3}\log\Big[\frac{1+2\alpha(\kappa^2)}{1-\alpha(\kappa^2)}\Big](w+w^{-1}),
\end{equation}
which can be checked to hold for any real parameter $\kappa$. With this illustrative example we conclude the Floquet interpretation of the time propagation.

\section{The hierarchy of Lax operators}\label{hierarchy}

In this appendix we introduce the hierarchy of the Lax operators, which allows one to write the Hilbert-Schmidt inner product of Faddeev-Volkov transfer operators in a more compact form. Recall the local operator basis \eref{operatorbasis}, consisting of elements ${\rm e}_{i,j}=u^iv^j$ with $i,j\in\mathbbm{Z}_m$.
Expanding the Lax operator \eref{lax} in this basis \eref{operatorbasis} according to $L(\lambda)=\sum_{i,j}L^{i,j}(\lambda)\otimes {\rm e}_{i,j}$, gives the following non-vanishing Lax components $L^{i,j}\in \End(\mathcal{V})$:
\begin{equation}
L^{1,0}(\lambda)=\ket{0}\bra{0},\quad
L^{m-1,0}(\lambda)=\ket{1}\bra{1},\quad
L^{0,1}(\lambda)=\lambda\ket{0}\bra{1},\quad
L^{0,m-1}(\lambda)=-\lambda\ket{1}\bra{0}.
\end{equation}
Let us denote the partial tensor product with respect to the physical Hilbert space $\mathcal{H}$ by $\otimes_p$, namely 
\begin{equation}
\big(A\otimes {\rm e}_{k,l}\big)\otimes_p\big(B\otimes {\rm e}_{i,j}\big)=AB\otimes {\rm e}_{i,j}\otimes {\rm e}_{k,l},\qquad A,B\in\End(\mathcal{V}).
\end{equation}
The transfer operator \eref{conserved} can now be rewritten as $T(\lambda)=\tr_{\mathcal{V}}\big({\rm\mathbf L}(\lambda)^{\otimes_p N}\big)$ with the \emph{staggered Lax operators}
\begin{equation}
{\rm\mathbf L}(\lambda)=L_2(\lambda/\kappa)L_1(\lambda \kappa)=\sum_{i,j,k,l}{\rm\mathbf L}^{[i,j,k,l]}(\lambda)\otimes {\rm e}_{i,j}\otimes {\rm e}_{k,l},\quad {\rm\mathbf L}^{[i,j,k,l]}(\lambda)=L_2^{k,l}(\lambda/\kappa)L_1^{i,j}(\lambda \kappa).
\end{equation}
By direct calculation, one can show that the only nonzero components of the staggered Lax operators are 
\eqnarray
\eqalign{
{\rm\mathbf L}^{[0,1,0,m\mn 1]}(\lambda)=-\lambda^2\ket{1}\bra{1},\quad{\rm\mathbf L}^{[0,1,1,0]}(\lambda)=\kappa\lambda\ket{0}\bra{1},\cr\cr
{\rm\mathbf L}^{[0,m\mn 1,0,1]}(\lambda)=-\lambda^2\ket{0}\bra{0},\quad{\rm\mathbf L}^{[0,m\mn 1,m\mn 1,0]}(\lambda)=-\kappa\lambda\ket{1}\bra{0}
\cr\cr
{\rm\mathbf L}^{[1,0,0,m\mn 1]}(\lambda)=-\frac{\lambda}{\kappa}\ket{1}\bra{0},\quad{\rm\mathbf L}^{[1,0,1,0]}(\lambda)=\ket{0}\bra{0},
\cr\cr
{\rm\mathbf L}^{[m\mn 1,0,0,1]}(\lambda)=\frac{\lambda}{\kappa}\ket{0}\bra{1},\quad{\rm\mathbf L}^{[m\mn 1,0,m\mn 1,0]}(\lambda)=\ket{1}\bra{1}.
}
\label{stag_comps}
\endeqnarray
Note, that they are independent of the order of the root of unity, $m$. Now we can continue to write
\eqnarray
T(\lambda_1)T(\lambda_2)=\tr{_{\mathcal{V}\otimes\mathcal{V}}\Big[\mathbbm{L}(\lambda_1,\lambda_2)^{\otimes_p N}\Big]}=\tr{_{\mathcal{V}\otimes\mathcal{V}}\Big[\Big(\sum_{ijkl}\mathbbm{L}^{[i,j,k,l]}(\lambda_1,\lambda_2)\otimes {\rm e}_{i,j}\otimes {\rm e}_{k,l}\Big)^{\otimes_p N}\Big]},\label{compactTT}\\[.75em]
\mathbbm{L}^{[i,j,k,l]}(\lambda_1,\lambda_2)=\sum_{i',j',k',l'}q^{(i'- i)j'+(k'- k)l'}\,{\rm\mathbf L}^{[i',j',k',l']}(\lambda_1)\otimes {\rm\mathbf L}^{[i\mn i',j\mn j',k\mn k',l\mn l']}(\lambda_2).\label{two_point}
\endeqnarray
In \eref{compactTT} we have implicitly defined the \emph{double Lax operator} $\mathbbm{L}$ with components $\mathbbm{L}^{[i,j,k,l]}$ in $\End(\mathcal{V}\otimes\mathcal{V})$. The compact formulas for the Hilbert-Schmidt inner product \eref{HS_1} and the auxiliary transfer matrix \eref{HS_2} now follow straightforwardly. The leading Lax component
\begin{equation}
\fl\mathbbm{L}_0(\lambda_1,\lambda_2)\equiv\mathbbm{L}^{[0,0,0,0]}(\lambda_1,\lambda_2)=\left(1+\lambda_1^2\,\lambda_2^2\right)\left(\ket{01}\bra{01}+\ket{10}\bra{10}\right)-(\kappa^2+\frac{1}{\kappa^2})\lambda_1\lambda_2\left(\ket{01}\bra{10}+\ket{10}\bra{01}\right),
\end{equation}
has two nontrivial eigenpairs $\mathbbm{L}_0(\lambda_1,\lambda_2)\ket{\psi_{s,t}}=\Lambda_{s,t}(\lambda_1,\lambda_2)\ket{\psi_{s,t}}$,
\eqnarray
&\Lambda_s(\lambda_1,\lambda_2)=1+(\kappa^2+\frac{1}{\kappa^2})\lambda_1\lambda_2+\lambda_1^2\lambda_2^2,\hspace{.35cm}\ket{\psi_s}=\frac{1}{\sqrt{2}}\left(\ket{01}-\ket{10}\right),\label{sin}\\
&\Lambda_t(\lambda_1,\lambda_2)=1-(\kappa^2+\frac{1}{\kappa^2})\lambda_1\lambda_2+\lambda_1^2\lambda_2^2,\hspace{.35cm}\ket{\psi_t}=\frac{1}{\sqrt{2}}\left(\ket{01}+\ket{10}\right).\label{tri}
\endeqnarray
The first one is referred to as the \emph{singlet eigenpair}, while the second one is referred to as the \emph{triplet eigenpair}. As shown in Section \ref{QL}, the quasilocality of our conservation laws stems from the factorization of these eigenvalues into the leading eigenvalue of the auxiliary transfer matrix \eref{HS_2}.

\section{The reduced auxiliary transfer matrix}\label{Matr} 

Recall that the factorizable auxiliary transfer matrix \eref{explicitfactorization} can be written as 
\begin{equation}
\mathbbm{T}(\lambda q^{\frac{1}{2}},\lambda q^{-\frac{1}{2}},\mu q^{\frac{1}{2}},\mu q^{-\frac{1}{2}})=\mathbbm{T}^{\rm (r)}(\lambda,\mu)\oplus 0.
\end{equation}
Here we state the explicit form of its nontrivial part, namely the reduced auxiliary transfer matrix $\mathbbm{T}^{\rm (r)}(\lambda,\mu)$:
\eqnarray
\fl
\tiny
\left(
\begin{array}{c c c}
\mu ^4 \overline{\lambda}^4+\left(q^2 +\frac{1}{q^2}+2\right) \mu ^2 \overline{\lambda}^2+1 
&\left(\kappa^2 +\frac{1}{\kappa^2}\right)\left(\overline{\lambda}^3 \mu ^3+\overline{\lambda} \mu\right)-\overline{\lambda} \mu^3-\overline{\lambda}^3 \mu
&\left(\kappa^2 +\frac{1}{\kappa^2}\right)\left(\overline{\lambda}^3 \mu ^3+\overline{\lambda} \mu\right)-\overline{\lambda} \mu^3-\overline{\lambda}^3 \mu
\\
\left(\kappa^2 +\frac{1}{\kappa^2}\right)\left(\overline{\lambda}^3 \mu ^3+\overline{\lambda} \mu\right)-q^2 \overline{\lambda} \mu ^3-\frac{\overline{\lambda}^3 \mu }{q^2}
&\overline{\lambda}^4 \mu ^4+2 \left(\overline{\lambda}^2+\mu ^2\right)^2+1 
&-\mu ^2\left(1+ \overline{\lambda}^4\right)\left(\kappa^2+\frac{1}{\kappa^2}\right)+2 \mu ^2 \overline{\lambda}^2
\\
\left(\kappa^2 +\frac{1}{\kappa^2}\right)\left(\frac{\overline{\lambda}^3 \mu ^3}{q^2}+q^2\overline{\lambda} \mu\right)-\overline{\lambda} \mu^3-\overline{\lambda}^3 \mu 
&-\mu ^2\left(1+\overline{\lambda}^4\right)\left(\kappa^2+\frac{1}{\kappa^2}\right)+\left(q^2+\frac{1}{q^2}\right)\mu ^2 \overline{\lambda}^2
&\overline{\lambda}^4+\overline{\lambda}^4 \mu ^4+\mu ^4+\left(q^2+\frac{1}{q^2}\right)\overline{\lambda}^2 \mu^2+1
\\ 
\left(\kappa^2 +\frac{1}{\kappa^2}\right)\left(\frac{\overline{\lambda}\mu}{q^2}+q^2\overline{\lambda}^3 \mu^3\right)-\overline{\lambda} \mu^3-\overline{\lambda}^3 \mu 
&-\overline{\lambda}^2\left(1+\mu^4\right)\left(\kappa^2+\frac{1}{\kappa^2}\right)+\left(q^2+\frac{1}{q^2}\right)\mu ^2 \overline{\lambda}^2
&\mu ^2 \overline{\lambda}^2 \left(\kappa^4+\frac{1}{\kappa^4}+q^2+\frac{1}{q^2}+2\right)
\\
\left(\kappa^2 +\frac{1}{\kappa^2}\right)\left(\overline{\lambda}^3 \mu ^3+\overline{\lambda} \mu\right)-q^2 \overline{\lambda}^3\mu-\frac{\overline{\lambda}\mu^3}{q^2}
&\mu ^2 \overline{\lambda}^2 \left(\kappa^4+4+\frac{1}{\kappa^4}\right)
&-\overline{\lambda}^2\left(1+\mu ^4\right)\left(\kappa^2 + \frac{1}{\kappa^2}\right)+2 \overline{\lambda}^2 \mu ^2 
\\
\mu ^2 \overline{\lambda}^2\left(\kappa^4+\frac{1}{\kappa^4}\right)+\left(q^2+\frac{1}{q^2}+2\right)\mu ^2 \overline{\lambda}^2
&\left(\kappa^2 +\frac{1}{\kappa^2}\right)\left(\overline{\lambda}^3 \mu ^3+\overline{\lambda} \mu\right)-\overline{\lambda} \mu^3-\overline{\lambda}^3 \mu 
&\left(\kappa^2 +\frac{1}{\kappa^2}\right)\left(\overline{\lambda}^3 \mu ^3+\overline{\lambda} \mu\right)-\overline{\lambda} \mu^3-\overline{\lambda}^3 \mu
\end{array}\right.
\normalsize
\cr\cr
\tiny
\fl
\left.\begin{array}{c c c}
\left(\kappa^2 +\frac{1}{\kappa^2}\right)\left(\overline{\lambda}^3 \mu ^3+\overline{\lambda} \mu\right)-\overline{\lambda} \mu^3-\overline{\lambda}^3 \mu
&\left(\kappa^2 +\frac{1}{\kappa^2}\right)\left(\overline{\lambda}^3 \mu ^3+\overline{\lambda} \mu\right)-\overline{\lambda} \mu^3-\overline{\lambda}^3 \mu
&\mu ^2 \overline{\lambda}^2\left(\kappa^4+\frac{1}{\kappa^4}\right)+\left(q^2+\frac{1}{q^2}+2\right)\mu ^2 \overline{\lambda}^2
\\
-\overline{\lambda}^2\left(1+\mu ^4\right)\left(\kappa^2 + \frac{1}{\kappa^2}\right)+2 \overline{\lambda}^2 \mu ^2  
&\mu ^2 \overline{\lambda}^2 \left(\kappa^4+4+\frac{1}{\kappa^4}\right)
&\left(\kappa^2 +\frac{1}{\kappa^2}\right)\left(\overline{\lambda}^3 \mu ^3+\overline{\lambda} \mu\right)-q^2 \overline{\lambda}^3\mu-\frac{\overline{\lambda}\mu^3}{q^2}
\\
\mu ^2 \overline{\lambda}^2 \left(\kappa^4+\frac{1}{\kappa^4}+q^2+\frac{1}{q^2}+2\right)
&-\overline{\lambda}^2\left(1+\mu^4\right)\left(\kappa^2+\frac{1}{\kappa^2}\right)+\left(q^2+\frac{1}{q^2}\right)\mu ^2 \overline{\lambda}^2 
&\left(\kappa^2 +\frac{1}{\kappa^2}\right)\left(\frac{\overline{\lambda}\mu}{q^2}+q^2\overline{\lambda}^3 \mu^3\right)-\overline{\lambda} \mu^3-\overline{\lambda}^3 \mu 
\\
\overline{\lambda}^4+\overline{\lambda}^4 \mu ^4+\mu ^4+\left(q^2+\frac{1}{q^2}\right)\overline{\lambda}^2 \mu^2+1  
&-\mu ^2\left(1+\overline{\lambda}^4\right)\left(\kappa^2+\frac{1}{\kappa^2}\right)+\left(q^2+\frac{1}{q^2}\right)\mu ^2 \overline{\lambda}^2 
&\left(\kappa^2 +\frac{1}{\kappa^2}\right)\left(\frac{\overline{\lambda}^3 \mu ^3}{q^2}+q^2\overline{\lambda} \mu\right)-\overline{\lambda} \mu^3-\overline{\lambda}^3 \mu  
\\
-\mu ^2\left(1+\overline{\lambda}^4\right)\left(\kappa^2+\frac{1}{\kappa^2}\right)+\left(q^2+\frac{1}{q^2}\right)\mu ^2 \overline{\lambda}^2 
&\overline{\lambda}^4 \mu ^4+2 \left(\overline{\lambda}^2+\mu ^2\right)^2+1
&\left(\kappa^2 +\frac{1}{\kappa^2}\right)\left(\overline{\lambda}^3 \mu ^3+\overline{\lambda} \mu\right)-q^2 \overline{\lambda} \mu ^3-\frac{\overline{\lambda}^3 \mu }{q^2}
\\
\left(\kappa^2 +\frac{1}{\kappa^2}\right)\left(\overline{\lambda}^3 \mu ^3+\overline{\lambda} \mu\right)-\overline{\lambda} \mu^3-\overline{\lambda}^3 \mu
&\left(\kappa^2 +\frac{1}{\kappa^2}\right)\left(\overline{\lambda}^3 \mu ^3+\overline{\lambda} \mu\right)-\overline{\lambda} \mu^3-\overline{\lambda}^3 \mu 
&\mu ^4 \overline{\lambda}^4+\left(q^2 +\frac{1}{q^2}+2\right) \mu ^2 \overline{\lambda}^2+1  
\end{array}
\right)\nonumber
\normalsize
\endeqnarray
For general spectral parameters $\lambda,\mu$ this matrix can be diagonalized, namely there exists an invertible square matrix $S(\lambda,\mu)$ such that $S(\lambda,\mu)^{-1}\mathbbm{T}^{\rm (r)}(\lambda,\mu)S(\lambda,\mu)$ is diagonal. There are four $q$-independent eigenvalues,
\eqnarray
&\tau_1(\lambda,\mu)=-\frac{(\kappa^2-\overline{\lambda}^2)(\kappa^2 \overline{\lambda}^2 -1) \left(\kappa^2+\mu ^2\right) \left(\kappa^2 \mu ^2+1\right)}{\kappa^4},\label{deg1}\\
&\tau_2(\lambda,\mu)=-\frac{(\kappa^2+\overline{\lambda} ^2) (\kappa^2 \overline{\lambda} ^2+1) (\kappa^2-\mu^2)(\kappa^2 \mu^2 -1)}{\kappa^4},\label{deg2}\\
&\tau_3(\lambda,\mu)=\overline{\lambda} ^4 \mu ^4-\frac{\left(\kappa^8+1\right) \overline{\lambda} ^2 \mu ^2}{\kappa^4}+1\\
&\tau(\lambda,\mu)=\frac{(\kappa^2+\overline{\lambda} ^2)(\kappa^2 \overline{\lambda} ^2+1) \left(\kappa^2+\mu ^2\right) \left(\kappa^2 \mu ^2+1\right)}{\kappa^4}.
\endeqnarray
We have been able to compute the remaining two eigenvalues analytically only in the simplest case, namely for the third root of unity, $q=\exp(i\frac{2}{3}\pi)$. For other roots of unity they can be computed numerically.

\end{document}